\documentclass[preprint,showpacs,preprintnumbers,amsmath,amssymb]{revtex4}

\usepackage{graphicx}
\usepackage{dcolumn}
\usepackage{bm}

\newcommand{\be}{\begin{equation}}
\newcommand{\ee}{\end{equation}}
\newcommand{\bea}{\begin{eqnarray}}
\newcommand{\nn}{\nonumber}
\newcommand{\eea}{\end{eqnarray}}

\newcommand{\p}{\partial}

\begin{document}

\title{Faithful transformation of quasi-isotropic to
Weyl-Papapetrou coordinates: \\A prerequisite to compare metrics.}

\author{G.~Pappas}
\author{T.~A.~Apostolatos}%
\affiliation{%
Section of Astrophysics, Astronomy, and Mechanics, \\Department of Physics,
University of Athens, \\Panepistimiopolis Zografos GR15783, Athens, Greece }%

\date{\today}

\begin{abstract}
We demonstrate how one should transform
correctly  quasi-isotropic coordinates
to Weyl-Papapetrou coordinates in order to compare the metric around a rotating star that has been
constructed numerically in the former coordinates with an
axially symmetric stationary metric that is given through an analytical form in
the latter coordinates. Since a stationary metric associated with an isolated object
that is built numerically
partly refers to a non-vacuum solution (interior of the
star) the transformation of its coordinates to Weyl-Papapetrou
coordinates, which are usually used to describe vacuum axisymmetric and stationary
solutions of Einstein equations, is not
straightforward in the non-vacuum region. If this point is \textit{not} taken into consideration, one
may end up to erroneous conclusions about how well a
specific analytical metric matches the metric around the star,
due to fallacious coordinate transformations.
\end{abstract}

\maketitle

\section{Introduction}
\label{sec:1}

Recently there has been quite a few attempts to describe the
 geometry around an astrophysical object, such as a rotating neutron
star, a strange star, or a black hole surrounded by an accretion
disk, through various types of analytical solutions of the vacuum
Einstein equations
\cite{BertiSterg,Stuart,Pachonetal,FrauKlein,Cabezetal}.
Numerous
people have produced a lot of families of such exact solutions
during the last decades following various generating techniques. The
metrics that correspond to these families of solutions are usually
parameterized by a few parameters. These parameters could then be
used to fit the specific characteristics of each type of central
object by appropriate tuning of the parameter values. Thus if the
matching between a large range of metrics that are constructed
numerically and a family of parametrized analytical metrics is quite
acceptable, the space-time neighborhood of a multi-parametric
central compact object could be represented quite accurately by this
family of exact solutions. One could then use these metrics to
explore analytically the behavior of orbits around such a central
object (compute the frequencies related with these orbits, find out
the innermost circular radius, etc.), or the other way around: probe
the physical characteristics of the central body itself by
exploiting the characteristics of the gravitational waves produced
by small objects orbiting around the central one.

Fortunately, nowadays, there is a large variety of analytical
solutions of vacuum Einstein equations, which could be used as
candidate metrics to describe well the exterior space-time of
axisymmetric astrophysical objects. Ernst \cite{ernst1} formulated
the Einstein equations in the case of axisymmetric stationary
space-times long time ago, while Manko et al. and Sibgatullin
\cite{twosoliton,manko,manko2,manko3,sib1,SibManko} have used
various analytical methods to produce such space-times parameterized
by various parameters that have a different physical context
depending on the type of each solution. Also, Neugebauer \cite{Neug}
have constructed a specific axisymmetric solution analogous to the
well known Schwarzschild and Kerr black hole solutions: it describes
a a rotating thin disk of dust. Among all these available solutions,
one has to choose a specific type of solution that relates better to
the particular astrophysical object, depending on the specific
physical characteristics one expects from such an object.

On the other hand various groups (see \cite{Sterg}, and for an
extended list of numerical schemes see \cite{Lrr}), that have
expertise in building relativistic models of astrophysical objects
with adjustable  physical characteristics, can construct the metric
inside and outside such objects by solving numerically the
full Einstein equations in stationary cases. Their numerical codes generate
metrics in tabulated form with numerical values that correspond to
the metric components at the grid points that have been assumed in their numerical
scheme.

In order to compare an analytical solution with a metric that has been constructed
numerically, one should make sure that the transformation of the
coordinates of the metrics to each other, if not the same,
is absolutely faithful. More specifically a problem arises when
one attempts to transform the quasi-isotropic coordinates, that are
usually used to describe a metric that has been constructed
numerically, both inside and outside the astrophysical object to
Weyl-Papapetrou coordinates, that are usually used in the analytical
expressions of the available stationary axisymmetric metrics which will be compared with the numerical one.
Although the transformation of coordinates is straightforward in
the vacuum region, the same type of transformation leads to
erroneous coordinates when it is used in the matter region. This
inconsistence is expected since an axisymmetric metric in
quasi-isotropic coordinates is described by four  metric functions,
while in the Weyl-Papapetrou coordinates there are only three
independent functions present in the usual stationary axisymmetric
metric (actually not all of them are independent, since in both
cases one of the metric functions is uniquely obtained, apart from a
constant, from the rest functions). The difference comes actually from the
fact that the former one is used for an axisymmetric stationary
solution of the general Einstein equations, while the latter one is
used for axisymmetric stationary \textit{vacuum} solutions of
Einstein equations. Of course when one deals with vacuum solutions
in quasi-isotropic coordinates the four metric functions are
interrelated to each other by an extra constraint. On the other hand
if one insists on using the Weyl-Papapetrou coordinates $\rho$ and
$z$ inside the matter, one could not anymore use them on equal basis
as in the usual vacuum axisymmetric stationary metric; instead one
more function $\Lambda(\rho,z)$ should be introduced to describe the
induced metric of the two-dimensional surface spanned be $\rho$ and
$z$, namely
\be
ds_{(\rho,z)}^2=\Omega^2\left(d\rho^2+\Lambda(\rho,z)~ dz^2\right)
\ee
c.f. \cite{Wald}.

In this paper we suggest that in order to translate the
former to the latter coordinates, the corresponding integration path that is
usually used to compute the $z-$coordinate (c.f.~Sec.~\ref{sec:3}) should avoid entering
the matter region. Moreover, since the exterior region is the one we are
interested to in order to exploit its characteristics to probe the
source of the gravitational field,  we are not really interested to know the
actual coordinates inside the matter.

The rest of the paper is organized as follows:
In Sec.~\ref{sec:2} we show the relation between the two sets of coordinates
and the recipe to compute the metric components in Weyl-Papapetrou coordinates
from the metric components in quasi-isotropic coordinates. In Sec.~\ref{sec:3}
we argue that the best path of integration to compute the $z-$coordinate is to
follow a meridian ($r=\textrm{const}$), starting from an equatorial point just outside
the star's surface, up to whatever angle $\theta$ and then move along the radial coordinate $r$
either inwards up to the surface of the star or outwards to infinity. We end up
this section by a practical formula for computing the $z-$coordinate. Furthermore we use this formula
to obtain the exact relation between the $z-$coordinate and the isotropic $r-$coordinate
in the Schwarzschild metric as a demonstration of of the proposed scheme.
In Sec.~\ref{sec:4} we give an estimate of the errors that arise from numerical
integration of $z$, and the corresponding errors that are induced in the metric components.
Once again we use the example of Schwarzschild space-time to measure these numerical
errors since in this very case we know exactly the $z$-coordinate and we can compare it with
the value of $z$ obtained numerically. We argue that the errors computed for the Schwarzschild case
are of the same order of magnitude ($\simeq 10^{-6}-10^{-7}$ in $g_{tt}$) as for any neutron star model obtained
by the numerical code of Stergioulas,
even for the most rapidly rotating ones, if the simple trapezoid rule is used in numerical integration.
Thus we conclude that any relative difference between a numerical and an analytical metric
of order higher than $10^{-6}$ should be attributed to a real non-matching of the metrics.
In the last section we summarize our conclusions and  show how much improved is
the comparison between the numerical and analytical metrics studied by
Berti and Stergioulas \cite{BertiSterg}, if the transformation of the
coordinates is done according to our proposed scheme.

All physical quantities used in this paper are in geometrized units ($G=c=1$).

\section{Coordinate transformation in vacuum.}
\label{sec:2}

The line element of an axisymmetric and stationary space-time in
quasi-isotropic coordinates assumes the following form:
\be
\label{CSTmetr}
ds^2=-e^{2\nu} dt^2+
      e^{2\psi} \left( d\phi-\omega dt       \right)^2+
      e^{2\mu}  \left( dr^2 + r^2 d\theta^2 \right),
\ee
where $\nu,\,\psi,\,\omega,\,\mu$ are functions of $r$
and $\theta$ alone. An alternative way to write this metric is by
replacing the $\psi$ function by a new function $B$ through
\be
e^{\psi}=Be^{-\nu}r\sin\theta
\ee
upon which the line element transforms to: 
\be
\label{CSTmetr2}
ds^2=-e^{2\nu}dt^2+
      B^2e^{-2\nu} r^2 \sin^2 \theta \left( d\phi-\omega dt \right)^2+
      e^{2\mu} \left( dr^2+ r^2 d\theta^2 \right).
\ee
We note once again that the metric written in any of the
above forms has full freedom to describe any stationary
axisymmetric solution of the full Einstein equations.\\
\indent On the other hand the Weyl-Papapetrou coordinates are very
good to describe any stationary axisymmetric solution of the
vacuum Einstein equations. The line element in these coordinates
is
\be
\label{Pap}
ds^2=-f\left(dt-w d\phi\right)^2+
      f^{-1}\left[ e^{2\gamma} \left( d\rho^2+dz^2 \right)+
      \rho^2 d\phi^2 \right],
\ee
where now the three functions $f,\,w,\,\gamma$
are functions of $\rho,\,z$ alone.

In order to transform the quasi-isotropic coordinates to
Weyl-Papapetrou coordinates (c.f.~\cite{BertiSterg}) one first defines the cylindrical
coordinates
\be
\label{cylind}
\varpi \equiv r\sin\theta, \qquad
\zeta \equiv r\cos\theta.
\ee
Then the Einstein field equations in vacuum ($R_t^t+R_{\phi}^{\phi}=0$)
imply (c.f.~\cite{Wald})
\be
\label{laplace}
{\partial^2 (\varpi B) \over \partial \varpi^2} +
{\partial^2 (\varpi B) \over \partial \zeta^2} =0.
\ee
Thus one could use a new coordinate
\be
\label{trho} \rho \equiv \varpi B= e^{(\nu+\psi)}
\ee
instead, which satisfies the two-dimensional Laplace
equation in the ($r-\theta$) surface. One could then define a
harmonic function that is conjugate to $\rho$, that is
\be
z = z (\varpi,\zeta),
\ee
which satisfies the Cauchy-Riemann conditions
\bea
\label{CauRie}
{\p z\over \p \varpi} &=& - \frac{\p \rho}{ \p \zeta} = -\varpi \frac{\p B}{\p \zeta}, \label{z1}\\
{\p z\over \p \zeta}&=&{\p \rho\over \p \varpi} = B+\varpi
\frac{\p B}{\p \varpi}. \label{z2}
\eea
At this point we should note that the above construction
of conjugate coordinates $\rho,\,z$ was feasible only in the
vacuum region. In the interior of the axisymmetric star, $\rho$
could still be defined as above but it does not anymore satisfy
the two-dimensional Laplace equation and thus the other coordinate
$z$ fails to be constructed as an harmonic conjugate of $\rho$.

Now in the vacuum region of space-time we can
integrate the Cauchy-Riemann conditions (\ref{CauRie}) with initial
value for z
\be
z(\varpi,\zeta=0)=0.
\ee
This corresponds to the equatorial plane of the star, or to be
more specific, to that part of the equatorial plane that lies
outside the star. The corresponding integration yields the value of $z-$coordinate at any
point outside the star. Although $z=0$ at the equatorial plane inside
the star as well, we cannot integrate these relations along a path that lies inside the
star since they do not hold in matter. Thus the path we choose to integrate the Cauchy-Riemann relations
should lie entirely in the vacuum region  up to the final point.

Besides transforming the $(r,\theta)$ ---or $(\varpi,\zeta)$--- coordinates to $(\rho,z)$
coordinates, one has to compute the new metric functions $(f,w,\gamma)$ from
the old metric functions $(\nu,\omega,\mu,B)$ that are supposed to be known
at the grid points used in the numerical code that generates them. By direct use of the Cauchy-Riemann relations (\ref{z1},\ref{z2}),
and subsequent substitution of the coordinates defined in relation (\ref{cylind})
we get
\bea
d \rho^2 + d z^2 &=&
\left[ \left( {\p \rho \over \p \varpi} \right)^2 +
       \left( {\p \rho \over \p \zeta }\right)^2 \right ]
(d \varpi^2 + d \zeta^2) \nn \\
                 &=&
\left[ \left( {\p \rho \over \p \varpi} \right)^2 +
       \left( {\p \rho \over \p \zeta }\right)^2 \right ]
(d r^2 + r^2 d \theta^2).
\eea
Finally by comparing the two metrics (\ref{CSTmetr2},\ref{Pap}), and
keeping the coordinates $(t,\phi)$ the same in both metrics, we
obtain
\bea
\label{MPa2}
f& = &e^{2 \nu}-\omega^2 \rho^2 e^{-2 \nu},\\
\label{MPa} w& = &-\frac{\omega \rho^2 e^{-2\nu}}{f},\\
e^{2 \gamma} & = &  f
\left[ \left( {\p \rho \over \p \varpi}\right)^2 +
       \left( {\p \rho \over \p \zeta }\right)^2 \right ]^{-1}
e^{2 \mu}.
\label{MPa3}
\eea
These new metric functions, that are computed from the old ones, along with
the new coordinates ---$\rho$ that is directly computed from the old coordinates, and
$z$ that is computed by integration along paths that lie entirely along vacuum regions---
complete the metric transformation in Weyl-Papapetrou coordinates.
Of course the integration of the Cauchy-Riemann relations in order to compute
the $z-$coordinate  cannot, in general, be performed analytically;
one should rely on some numerical scheme to integrate the corresponding
relations. This technical issue will be addressed in the following Section.

\section{Transformation to Weyl-Papapetrou coordinates.}
\label{sec:3}

In this section we will present a practical recipe that one could follow
to integrate the Cauchy-Riemann relations in order to obtain the Weyl-Papapetrou $z-$coordinate.
As mentioned in the previous section, the other conjugate Weyl-Papapetrou coordinate, $\rho$,
is directly obtained from the metric function $B$ and the $\varpi-$coordinate  (cf.~Eq.~(\ref{trho})).
Through the Cauchy-Riemann relations (Eqs.~(\ref{z1},\ref{z2})) it is easy to verify that
\bea
\frac{\p z}{\p r}     &=& \cos\theta B + \sin\theta \frac{\p B}{\p \theta}, \label{dzdr}\\
\frac{\p z}{\p \theta}&=& -r^2 \sin\theta \left( \frac{\p B}{\p r}+\frac{B}{r} \right).
\eea
Moreover since $\mu\equiv\cos\theta$ (not to be confused with the corresponding metric function in
quasi-isotropic coordinates) and $x\equiv r/(r+r_e)$, where $r_e$ is the equatorial radius of the star,
are usually used in numerical schemes
to produce numerical values of the metric functions in a specific grid that is more uniformly
distributed in azimuthal angles and covers better the whole space up to infinity, the above derivatives
could straightforwardly be transformed to
\bea
\frac{\p z}{\p x}     &=& \frac{r_e}{(1-x)^2} \left(\mu B + (\mu^2-1) \frac{\p B}{\p \mu} \right), \label{dzdx} \\
\frac{\p z}{\p \mu}   &=& r_e  \left( x^2 \frac{\p B}{\p x}+B
\frac{x}{1-x} \right).\label{dzdmu}
 \eea

Equipped with these expressions, we may now choose a suitable path
to integrate them in order to assume the numerical values of the new
$z-$coordinate. In realistic rotating stars the equatorial radius of
the star $r_e$ is the maximum value of $r_s(\theta)$, that is the function
that describes the shape of the surface of the star. Thus we could
simply start from an equatorial point ($\theta=\pi/2$) just outside
the surface of the star where $z=0$, and follow the grid points along the meridian $x=x_0$
($r =r_0 \cong r_e$), until we reach  the axis of symmetry ($\theta=0$). Upon reaching whatever
intermediate angle $\theta$, we could then move radially (along $r$,
or $x$) ---either outwards, or inwards up to the surface--- to
obtain the numerical value of $z$ at every  grid point in the vacuum
region where the metric is known. We should stress at this point
that along the meridian of constant $x=x_0$ the optimum integration
is achieved by using expression
(\ref{dzdmu}), while along the radial direction at constant $\theta$
(or $\mu$) the optimum integration is achieved by using
expression (\ref{dzdr}) instead of (\ref{dzdx}); this difference in effectiveness of the two
relations with respect to radial changes of $z-$coordinate arises because the
corresponding integrant is well behaved in the former
case as $r\rightarrow \infty$, in contrast to what's happening in the latter case as $x
\rightarrow 1$. Thus the $z$ value at the grid point $(x,\mu)$ is
the outcome of the integral
\bea z(r,\mu) &=& r_e \left(x_0^2
\int_0^\mu   \frac{\p B}{\p x} d\mu'+
                       \frac{x_0}{1-x_0} \int_0^\mu B
                       d\mu' \right) \nn \\ &+&
\left(                 \mu \int_{r_0}^r B dr'+
                       (\mu^2-1)         \int_{r_0}^r \frac{\p B}{\p \mu} dr'
                    \right).
\label{integrz}
\eea
The above expression simplifies much if the
meridian path, followed initially, corresponds to exactly the
equatorial radius, since then $x_0=1/2$ ($r_0=r_e$). Practically
though, since the numerical computation of the derivative  $\p
B/\p x$ needs at least two neighboring grid points lying in the
vacuum region, the meridian path should  correspond to the $r$
value of the next after the first grid point  lying outside the
equator of the star. Finally the last integral term in the
expression above could be omitted if we seek to compute the $z$
values at the grid points along the axis of symmetry since there
$\mu=1$.

Of course one could choose any other path starting from the equator where $z=0$ to reach the final grid point, but
since the integration will be carried  numerically it is better to choose a path that minimizes the numerical errors.
In the next section we will show why the path suggested above is expected to be efficient with respect to
numerical errors and we will give an estimate of the error magnitude. Heuristically, the basic argument in favor of this path is the
fact that if we follow to move along
another meridian $x=\textrm{const}$ which is far outside the surface
of the star, the error in the
numerical computation of $\p B/\p x$
will be much greater, since $r(x)$ has a rapidly increasing derivative as $x
\rightarrow 1$. This numerical error will then follow as a
systematic error in all $z$ values when the integration along $x$
is computed next.

We will end this section by demonstrating this coordinate transformation by a very simple example
where the integrations could be performed analytically; namely the Schwarzschild metric. In the Appendix
one could find the form of the line element of this metric in isotropic coordinates, and indirectly in Weyl-Papapetrou
coordinates. The $B$ function for the Schwarzschild metric is
\be
B=1-\frac{M^2}{4 r^2}=1-\frac{M^2}{4 r_e^2}\left( \frac{1-x}{x} \right)^2.
\ee
Thus, only the $\p B/\p x$ and $B$ parts will survive in the integrands of
Eq.~(\ref{integrz}). For example along the positive part of the $z$-axis ($\mu=1$) the value of $z$
is easily computed to
\be
z(r,\mu=1)= r_e \left( \frac{1}{4}  \left.\frac{\p B}{\p x}\right|_{x_0=1/2} +
B |_{x_0=1/2} \right) + \int_{r_e}^{r} B(r') dr'.
\ee
By a few simple substitutions back and forth between the coordinate $r$ and the compactified coordinate $x$
we get at the end
\be
z(r,\mu=1)=r \left( 1+ \frac{M^2}{4 r^2} \right) .
\ee
This is the exact relation between the two sets of coordinates  (see Appendix).
In the next section we will use once again the Schwarzschild example to get
a first estimate of the errors arising from numerical implementation of the recipe described above.

\section{Estimate of numerical errors}
\label{sec:4}

In order to check whether a metric given in analytical form describes faithfully the
metric of a physically realistic configuration which has been constructed through numerical schemes
and thus its components are given in tabulated form
at the grid points where the metric has been computed, one should check how well the two metrics coincide
at these specific grid points. In the usual case, where the two types of metrics refer to different
kind of coordinates, one should first transform one set of coordinates to the other set so as to compare
the two metrics at the same points and then decide about the faithfulness of the specific analytical metric.
However the transformation of coordinates involve numerical errors since, in practice, the integration
associated with the computation of the new coordinates will in general  be performed through numerical integration.
Especially if we have the case presented in the previous sections where the isotropic coordinates of the numerical
metric have to be transformed to Weyl-Papapetrou coordinates of the analytical metric, the integration
will be based on the numerical metric that is given in discrete form and thus the implemented errors
could not be optimized further than a minimum value related to the number of the grid points.
Therefore one should first have an estimate of these errors before evaluating the pure differences between the two
metrics. Hence if the differences computed between the two metrics are of the order of the numerical errors
induced in the metric by the numerical transformation of coordinates, we could not assess any
countable difference between the two metrics.

An alternative way to tackle the problem of comparing the two types of metric, that although it implies numerical errors these could in principle
be minimized at will, is to transform the coordinates in which the analytical metric is expressed to the
other set of coordinates. Since the former metric is analytically known, the corresponding integrations used to compute
the new coordinates could be performed at whatever level of accuracy one desires. In that case
the difference between the metrics at the same grid points at which the numerical metric is known
is true and does not correlate at all with the computation of the new coordinates. In the case considered in
this paper this kind of transformation of coordinates is not that simple as the inverse one. Actually there is no
direct way to compute the isotropic coordinates ($r,\theta$) from the Weyl-Papapetrou coordinates ($\rho,z$)
as well as  the metric expressed in the latter ones. Therefore we will resort in the first method, that was
described in detail in the previous section,  to transform coordinates.
We will show though that the errors induced from numerical integrations, at least for the number of grid points
used in a specific physical example, is not much larger than the accuracy at which the numerical metric is known.
This fact suggests that there is no harm in using this method to transform coordinates and then compare metrics.

Let's say then that we want to compare two metrics, a numerical one $g^{(N)}_{\alpha \beta}(r,\theta)$, that corresponds to
a rotating neutron star with specific internal physical characteristics, and an analytical one $g^{(A)}_{\alpha \beta}(\rho,z)$
which we believe, or simply want to check if, it describes quite well the former numerical one.
As is shown in the corresponding variables of the two functions the two metrics are assumed functions of
different coordinates. We decide to compare the two metrics along the axis of symmetry of their axially symmetric space-time.
Anyway the comparison along the axis of symmetry seems to be more demanding for the numerical transformation of coordinates
since the integration path from the
equator to the $z$-axis is longer than to any other angle, and therefore it imposes larger error contribution
from the first two integrals of Eq.~(\ref{integrz}). On the other hand the last two integral terms in  (\ref{integrz})
are competitive to each other for various values of the angle parameter $\mu$. Although we cannot  draw a general
rule about the angle parameter $\mu$ at which these two terms assume the highest total value, we have seen
in practice that the $z$-axis is really the most heavily infected  direction from numerical errors in computing
the $z$-coordinate.

\begin{figure}

\includegraphics[width=.7\textwidth]{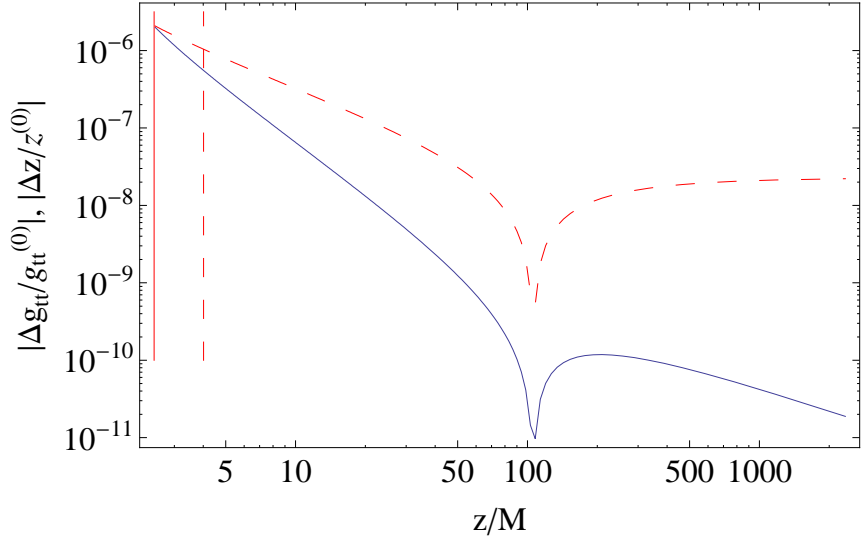}
\caption{The plot shows in a log-log plot
the relative difference between the exact $z$ value ($z^{(0)}$) and the $z$-coordinate produced through
numerical integration ($z^{(N)}$) at fixed grid points (upper thick dashed curve). The model used to estimate
the numerical error is a Schwarzschild metric that has the same gravitational mass
with the most rapidly rotating model of neutron star in Table 3 of \cite{BertiSterg}; namely $M=1.864 M_\odot$.
The radius used to describe the radial distances in the grid of our numerical integrations of $z$ is $r_e=10.755\textrm{~km}$,
which corresponds to the equatorial radius of the same rotating neutron star model.
At the same diagram we have also plotted the relative difference between
the exact metric component $g_{tt}^{(0)}$ and the one computed from transforming ($r,\theta$) to ($z,\rho$)
coordinates (lower thin solid curve). It is clear that the errors are at most of the order of $\sim 10^{-6}$ in
$g_{tt}$. Therefore any difference between a numerical metric and an analytical one should be of
order at least $10^{-6}$ to be considered true. The dashed vertical line marks the position on the $z$-axis
that one reaches by integrating along the meridian of constant radius $r=r_0$. The intersection of this line with the two
curves depicts the  systematic error induced in $z$ (and correspondingly in $g_{tt}$)
by replacing the $\p B/\p x$ term in integration along the meridian by its
numerical value. The errors at other locations along the $z$-axis are caused by numerical
$r$-integration following the trapezoid rule. The
deep wells in the plots are due to opposite signs of the errors accumulated along $\mu-$
and along $r-$integration in the region $r>r_0$
that end up nullifying  the difference between $z^{(0)}$ and $z^{(N)}$ at some point. Finally the vertical solid line
on the leftmost part of the plot marks the
polar surface location of the rotating neutron star  model cited above in order to indicate the minimum
$z-$value at which there is any physical meaning to look for numerical errors due to coordinate transformation.}
\protect\label{figure1}

\end{figure}

In Figure \ref{figure1} we have plotted the logarithmic relative
error in numerical computation of the $z$-coordinate in
Schwarzschild metric along the $z$-axis (the metric used corresponds
to a spherically symmetric star that has the same mass $M$ and
equatorial coordinate radius $r_e$ as the most rapidly rotating
neutron star model for which we want to estimate the errors caused
in $g_{tt}$). The number of grid points assumed are the ones used in
all numerical metrics that have been constructed for various models
of rotating neutron stars by Berti and Stergioulas \cite{BertiSterg}
and a simple trapezoid method of integration is used. Even though
this example could not be considered suitable to check the errors in
a realistic rotating neutron star case (due to spherical symmetry of
Schwarzschild metric $B$ is not a function of $\mu$, hence the
computation of the first two integrals in Eq.~(\ref{integrz}) do not
contribute any error), it gives at least a minimum estimate of the
order of magnitude of errors. In the spherically symmetric case the
errors arise from the numerically estimated value of $\p B/\p x$ in
the first integral of Eq.~(\ref{integrz}), and the numerical
computation of the third integral (the last integral is zero in
every case along the $z$-axis). Although the $z$-coordinate that we
compute numerically gets shifted more and more dramatically as $z$
increases due to cumulative errors along the integration path, the
error induced in the metric itself (e.g.~the $g_{tt}$ component)
does not increases so much with $z$. This is expected since the
metric becomes less sensitive to $z$ as we recede from the neutron
star. Consequently since the metrics themselves are the ones that we
want to compare, the cumulative error in $z$ at large values of $z$
is not disturbing. In Figure \ref{figure1} we have plotted  the
corresponding logarithmic relative error in $g_{tt}$, that is due
only to erroneous numerical integration of $z$ coordinate, along the
$z$-axis. As shown in the plot the relative error does not even
exceed the $\sim 10^{-6}$, which is actually just about one order of
magnitude higher than the level of accuracy of the numerical metrics
produced by the numerical code of Stergioulas \cite{Sterg}. Thus we
conclude that if the size of the rest of errors which are coming
from the $\mu-$dependent terms of a realistic numerical model do not
exceed the errors arising in the simple Schwarzschild case there is
no need to worry about any numerical errors induced to metrics
caused by transformation of coordinates.

In the remaining part of this section we will argue that this is
exactly the case with a generic numerical metric, as long as the
number of grid points in the vacuum region is sufficiently high
(of the order of $300 \times 300$ for the the $\mu$ and the $x$
coordinates in the vacuum region). To show this we shall appeal to the fastest rotating
neutron star model with EOS FPS used in \cite{BertiSterg} (the model that shows
up in the last line
of the sequence with $M_B=2.105M_{\bigodot}$ of Table 3 of the corresponding paper) to estimate the magnitude
of the error in computing the $z$-coordinate. Of course in the
case of a metric that is given in tabulated form we do not have a
true value of the $z$ coordinate to compare with, as in the simple
example analyzed previously. On the other hand by a simple plot of
$B(x_0,\mu)$ and $(\p B(x_0,\mu)/\p x)_{N}$ as a function of $\mu$
(where $x_0$ denotes the value of $x$-coordinate used for the
first part of the integration path; namely the path along the meridian
just outside the equator of the star, and the subscript $_{N}$
refers to the fact that the derivative is computed numerically),
we conclude that both these functions are quite constant along the
same meridian ($B$ and $(\p B/\p x)_{N}$ do not change by
more than $0.1 \%$ and  $25 \%$, respectively, over the whole
range of $\mu$), and thus the error in computing the first two
integrals in (\ref{integrz}), even by the simple trapezoid rule is
of the order of
\be
(\Delta z)_{\mu-\textrm{integration}} =
\frac{1}{h} \frac{h^3}{12} f''(\xi) < \frac{h^2}{10} \times
\max_{\mu}{f''(\mu)},
\ee
where $h$ denotes the step size of $\mu$
used in numerical integration ($1/h$ is the total number of
steps), while $\xi$ is some value of $\mu$ in the interval
$[0,1]$. We could form an upper value of this error by simplifying
$1/12$ to $1/10$ and using the maximum value of $f''(\mu)$ instead
of $f''(\xi)$. It is easy to verify that this error in numerical
integration over $\mu$ is of the order of only $\Delta z/M \simeq
3 \times 10^{-7}$. This is systematically lower, or at most of the same order of magnitude, than the
errors related to $x$-integration and numerical computation of
$\p B/\p x$, that were estimated previously by means of the
Schwarzschild example (c.f.~Fig.~\ref{figure1}). Therefore, the
plot of Fig.~\ref{figure1} summarizes quite well the overall
numerical errors in computing the values of $z$ and the
corresponding errors induced in computing the $g_{tt}$ component
of a numerically constructed metric.

We thus conclude that relative differences of metrics at the level of $10^{-6}$ and higher are true differences in metrics,
and only these should be taken seriously into consideration when a proposed analytical metric is used as a faithful representation
of a metric that is constructed numerically. The rest discrepancies between metrics could be easily attributed
to inexact transformation of the coordinates.

\section{Conclusions}
\label{sec:5}

In this paper we have noted that when comparing numerical metrics with analytical metrics, where the two
metrics are expressed in different coordinates, one should be very careful in transforming coordinates.
Especially since the Weyl-Papapetrou coordinates are usually used in vacuum Einstein stationary axisymmetric
solutions, one should avoid the interior of a
star in the  integration path used to transform the initial quasi-isotropic coordinates, or else one results in
erroneous comparison between metrics. Since  the Cauchy-Riemann relations for the Weyl-Papapetrou coordinates
do not hold in matter,
integration of these relations along a path passing through the interior of the star
results in a systematic shift of the $z$-coordinates, and consequently one ends up comparing two
metrics at completely different points.

\begin{figure}

\includegraphics[width=.7\textwidth]{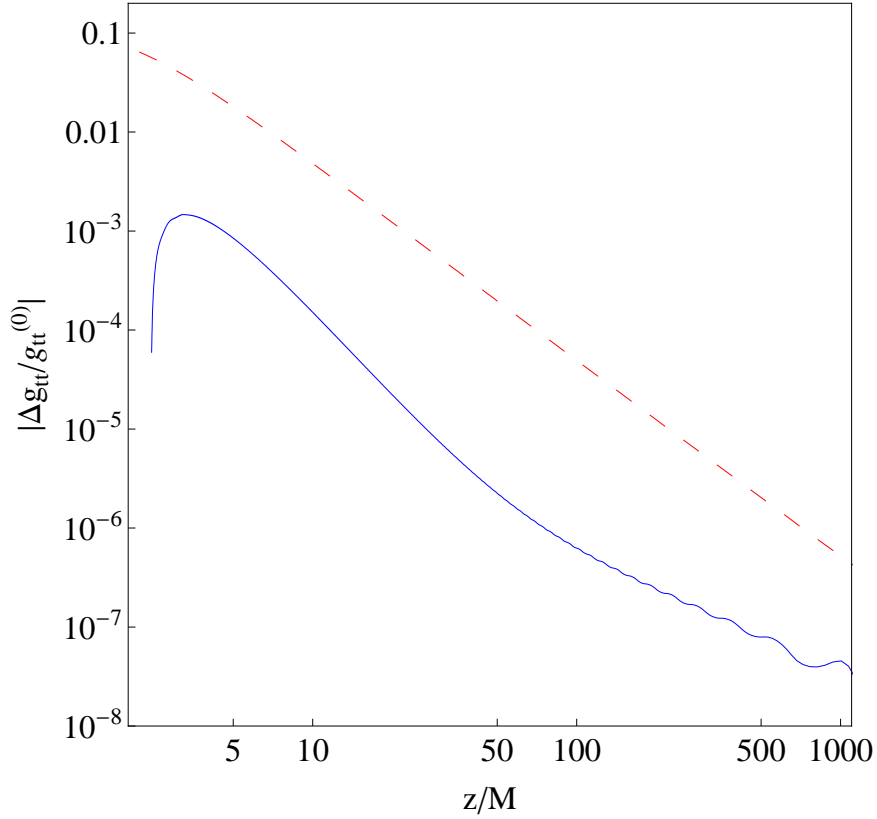}
\caption{This is a replot of Figure 6 of \cite{BertiSterg} (upper
curve) with somewhat altered dimensions along the horizontal axis
(it's dimensionless here). The same difference between exactly the
same two metrics is plotted for contrast (lower curve) when the
transformation of coordinates is done according to our proposed
method. The similarity between the analytic metric and the one at
the exterior of the rapidly rotating neutron star model is
manifestly much better than what was considered initially. The odd
behavior of the lower curve near the surface of the star
demonstrates a possible (probably accidental) crossing approach
between the two metrics. The apparent oscillatory behavior of the
rightmost part of the lower curve is due to the fact that the
difference between the two metrics have approached the level of
$10^{-8}-10^{-7}$. This is actually the level of accuracy of the
numerical method that have generated the numerical metric an thus
spurious numerical information may have been introduced in the
metric at this level.} \protect\label{figure2}

\end{figure}

By a thorough error analysis of the most stringent case (a maximally rotating neutron star model),
we have  concluded that when comparing a numerical metric with an analytical one, through
faithful transformation of coordinates, any relative differences
that exceed the level of $10^{-6}$ should be  considered real and not an artifact
of the numerical transformation of coordinates, at least for metrics constructed on grids
with grid size of at most that used by Stergioulas \cite{Sterg}.
In Fig.~\ref{figure2} we have plotted once again the
Figure 6 of \cite{BertiSterg}, but now we have followed the path in the vacuum region just outside the star
(the one described in Sec.~\ref{sec:3}) to integrate the coordinate $z$ in order to transform
the numerical metric components and compare them with the metric described by the solution of Manko et al. \cite{manko}.
By direct inspection we find out that the matching between the two metrics is even better than
what is inferred by Berti and Stergioulas  \cite{BertiSterg}. The right transformation of the $z$-coordinate
leads to about two orders of magnitude better matching between the two metrics than the one presented in
\cite{BertiSterg}, even right at the surface of the star (leftmost part of the diagram). This renders the
metric introduced by Manko et al.  a very good candidate (even better than what it was first considered)
to describe the space-time around a rotating neutron star.

In a forthcoming paper we examine another similar candidate
analytical metric (the one described in \cite{twosoliton,
PapSotiri}) to describe the space-time around any kind of neutron
star, either rotating or not. Remember that the analytical metric
used by Berti and Stergioulas had the disadvantage that it could
not be adjusted to describe very slow rotating stars, since the
corresponding metric could not be made to erase simultaneously
both its quadrupole moment and its spin.

\section*{Aknowledjements}

We would like to thank Nikolaos Stergioulas for many useful
discussions and for providing us with access to his numerical code.
G. Pappas would also like to thank the Aristotle University of
Thessaloniki and especially the Astronomy Laboratory of the Physics
Department for their hospitality. This work was supported partly by
the ``PYTHAGORAS'' research funding program Grant No 70/3/7396, and
partly by the research funding program ``Kapodistrias'' with Grant
No 70/4/7672.

\section*{Appendix: The Schwarzschild metric in isotropic and Weyl-Papapetrou coordinates}

Here we write down the Schwarzschild metric in both isotropic and Weyl-Papapetrou coordinates
in order to use it to measure the errors induced when $(r,\theta)-$coordinates are numerically
transformed to  $(\rho,z)-$coordinates. Since in this example the exact metric is given as an analytical function
in both sets of coordinates, we could infer what fraction of discrepancy between a numerical metric transformed into
Weyl-Papapetrou coordinates and an analytical metric is caused by numerical errors in
transforming the coordinates (see Sec.~\ref{sec:4}).

The spherically symmetric metric in Schwarzschild
coordinates $(t,\tilde{r},\theta,\phi)$ has the form:
\be
ds^2=-\left(1-\frac{2 M}{\tilde{r}}\right) dt^2+
\left(1-\frac{2 M }{\tilde{r}}\right)^{-1} d\tilde{r}^2+
\tilde{r}^2(d\theta^2+\sin^2 \theta d\phi^2).
\ee
The same metric in isotropic coordinates $(t,r,\theta,\phi)$  assumes the
form:
\be
ds^2=-e^{2\nu} dt^2+e^{2\lambda}\left(dr^2+r^2d\theta^2+r^2\sin^2\theta ~d\phi^2\right),
\ee
where
\be
e^{2\nu}=\frac{\left(1-\frac{M }{2 r}\right)^2}{\left(1+\frac{M}{2 r}\right)^2}\;,\;
e^{2\lambda}=\left(1+\frac{M }{2 r}\right)^4.
\label{isoSchw}
\ee
The two radial coordinates, $\tilde{r},r$ are related to each other by
\be
\tilde{r}=r \left(1+\frac{M}{2 r} \right)^2.
\ee
From the form of the metric in isotropic coordinates (\ref{isoSchw}) one gets directly the $B$ function:
\be
B= 1 - \frac{M^2}{4 r^2}.
\ee
Finally, the same metric in Weyl-Papapetrou coordinates gets a  complicated form,
which could be obtained by writing down the inverse transformation of
\bea
\rho &=& r \sin\theta \left(1 - \frac{M^2}{4 r^2} \right) \label{rhoSchw}\\
z &=& r \cos\theta \left(1 + \frac{M^2}{4 r^2} \right) \label{zSchw}
\eea
and then constructing the metric components according to
the recipe given in relations (\ref{MPa2},\ref{MPa},\ref{MPa3}).
Especially on the $z$-axis ($\theta=0$), Eq.~(\ref{zSchw}) could be easily
solved with respect to $r$:
\be
r=\frac{z+\sqrt{z^2-M^2}}{2}.
\ee
This function of $z$ could then be used to obtain the exact value of all metric components as functions of $z$-coordinate.
For example the exact value of $g_{tt}$ component, which we use in the paper to compare metrics, yields the following simple expression
in Weyl-Papapetrou
\be
g_{tt}=-e^{2 \nu}=\frac{M-z}{M+z}.
\ee

\end{document}